\documentclass[aip,jcp,nofootinbib,reprint]{revtex4-1}

\usepackage{graphicx}
\usepackage{color}
\usepackage{dcolumn}
\usepackage{bm}
\usepackage{textcomp}
\usepackage{multirow}
\usepackage{array}
\usepackage{amssymb,amsfonts,amsmath}

\begin{document}

\title{THz time-domain spectroscopic investigations of thin films}

\author{A. Taschin$^{1}$, P. Bartolini$^{1}$,J. Tasseva$^{2,3}$, and R. Torre$^{4,}$ }
\email{torre@lens.unifi.it}

\affiliation{
$^1$European lab. for Non-Linear Spectroscopy (LENS), Univ. di Firenze, via N. Carrara 1, I-50019 Sesto Fiorentino, Firenze, Italy.\\
$^2$INFN, Istituto Nazionale di Fisica Nucleare, Sez. di Napoli, Complesso Univ. di M. S. Angelo, Ed. 6, Via Cintia, 80126 Napoli\\
$^3$CNR-INO, Istituto Nazionale di Ottica, Via Campi Flegrei 34, 80078 Pozzuoli, Italy.\\
$^4$Dip. di Fisica e Astronomia, Univ. di Firenze, via Sansone 1, I-50019 Sesto Fiorentino, Firenze, Italy.}

\begin{abstract}
THz time domain spectroscopy is a powerful technique enabling the investigation of different materials in the far-infrared frequency range. Even if nowadays this technique is well established, its application to very thin films remains particularly difficult. We investigated the utilization of THz spectroscopy on samples of micrometric thickness with the aim to disentangle multiple reflections and to measure with high accuracy the absolute values of the material parameters. We implemented an experimental and data analysis procedure that can be applied to free-standing single-layers or multi-layers samples. 

Specifically, we report on the experimental investigation by THz time domain spectroscopy of two samples: a test sample made of two layers of known thickness and materials; and a second sample, that is of a great interest for cultural heritage studies, made of a thin film of ink layered on a thicker support. Moreover, we describe in details the data analysis and fitting procedures needed to extract the material parameters from the experimental results. 
\end{abstract}

%
%
\maketitle

\section{Introduction}

The realization of new coherent radiation sources in the THz frequencies boosts dramatically the development of innovative spectroscopic techniques~\cite{Lee_09}. These spectroscopies are non-invasive methods that provide complementary information to traditional analytical tools. THz radiation introduces lower risks in terms of sample preservation so it must be considered as a particularly suitable probe for fragile/sensible samples. The potential to provide non-destructive information in the cultural heritage field has been demonstrated in a series of recent studies~\cite{Fukunaga_08,Abraham_09,Labaune_10,Seco_13,Bardon_13,Walker_13,Krugener_15,Jackson_15}.

Nowadays, THz spectroscopic investigations are supported by a series of commercial off-the-shelf systems. Typically, these systems are not open to a full control of the experimental parameters, so the lab customized set-ups prove more flexible and adaptable to a specific application. 

In the last years, the THz-Time Domain Spectroscopy (THz-TDS) has been recognized as a leading tool to measure the transmission parameters of complex materials. This is a spectroscopic method based on THz pulsed radiation generated by down-conversion of ultrafast optical laser pulses. Even if the THz-TDS is nowadays a well established technique, its application to samples characterized by a complex structure remains an open problem.

In this work, we explored the potentiality of a specific THz-TDS experimental apparatus, based on a table top set-up, in order to investigate samples formed by multiple layers structures and characterized by micrometric thickness. We implemented a quite efficient data analysis and fitting procedure that enable the extraction of the material optical parameters (i.e. absorption and index of refraction) with absolute values and the measurement of the layer thickness down to tens of micrometers. In this paper we report on the numerical algorithm that defines the iterative fitting process. 
We applied the experimental and data analysis methods to the specific problem of measuring the THz spectral features of thin bi-layer samples: a test sample made of a plastic layer on a Teflon substrate and a prototype sample of interest for application in artworks studies made of a thin ink layer deposited on a polyethylene support.

\section{THz time-domain spectroscopy set-up}

\begin{figure*}[htb]
\centering
\includegraphics[width=0.8\textwidth]{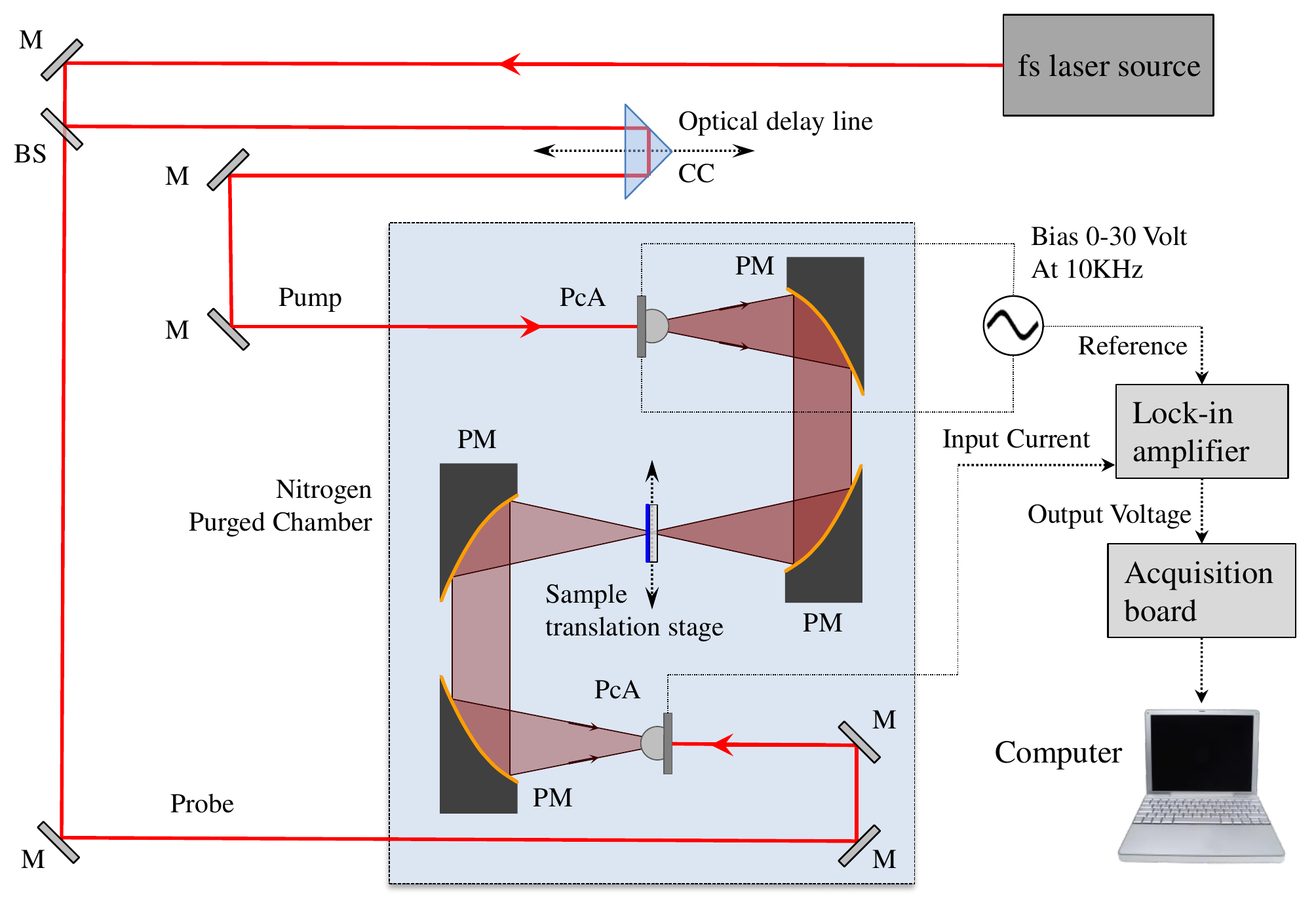}
\caption{
We report a sketch of the experimental set-up utilized to realize the THz time-domain spectroscopy, the sample investigation is performed in THz transmission configuration. The labelled elements are: M – mirror, BS – beam splitter, CC – corner cube, PcA – photoconductive antenna, PM – parabolic mirror.}
\label{setup}
\end{figure*}

The data presented in the current paper have been measured by a home-made THz-TDS system in transmission configuration, which enables us to explore the frequency range 0.1 - 4 THz. In Figure~\ref{setup} we show a sketch of our THz-TDS set-up. The THz pulses are produced by exciting a Low-temperature GaAs photoconductive antenna (PcA)~\citep{Lee_09} with optical laser pulses, at $\lambda=780~nm$, pulse duration of around $120~fs$ and a repetition rate of $100~MHz$ (produced by a T-light 780 nm fiber laser from MenloSystems). PcA is biased with sinusoidal voltage of $0-30~Volt$ at a frequency of $10~KHz$. 
%
%

The free carriers, generated by the laser pulse focused on the dipole gap, are accelerated by the bias field and quickly recombine producing a short current pulse. This, in turn, generates a short burst of electromagnetic radiation with a broad spectrum in the THz region. The emitted THz field is efficiently extracted from the chip by a hemispherical silicon lens, and then collimated and focused on the sample by two parabolic off-axis mirrors (PMs). The THz pulse transmitted through the sample is again collimated and focused, by another couple of PMs, on a second PcA. Another hemispherical silicon lens optimizes the coupling between the THz field and the dipole of the detection antenna. The THz field, in this case, acts as a bias, which accelerates the free carriers produced by a second optical pulse, the probe, focused again on the gap dipole. Thanks to the shorter temporal width of the laser pulses compared to that of the THz ones, the probe acts as a current gate and the amplitude of the photocurrent is directly related to the amplitude of the THz field. The whole time evolution of the electric field of the THz pulse is then obtained recording the photocurrent amplitude at varying the time delay between the pump and probe pulses. The detection current is amplified by a lock-in amplifier, locked at the bias frequency of the source antenna, and digitalized by an acquisition board. A home-made software acquires the processed signal together with the reading of the delay line encoder and retraces the final time dependent THz field.
The whole THz set-up is enclosed in a nitrogen purged chamber for removing the water vapour contribution present at the THz frequencies spanned by the experiment.

As reported in the next section, the optical properties of a material can be calculated measuring the amplitude, phase and time delay modifications that the THz pulse undergoes by crossing the sample. This is obtained by calculating the ratio between the Fourier transforms of the THz pulse which has crossed the sample and the one of the reference pulse obtained without any sample; this ratio is referred to as a transfer function. In order to improve the data quality and reduce the effects of external perturbations during the acquisition, the sample is mounted on a motorized translation stage, for moving the sample outside and inside the THz path. We performed several scans for the two configurations alternating the position of sample stage from reference to sample, thus for every sample scan we took a reference scan. Then we performed the average of the transfer functions obtained from each couple of sample and reference signals. Each single scan is acquired for $300$ s at a rate of $10~KHz$ with a continuous motion of the probe delay line at a velocity of $0.5~mm/s$.

\section{Material parameters extraction from experimental data}

The ratio between the Fourier transform of the THz pulse, transmitted after the sample, $E_t \left( \omega\right)$, and the one of the reference pulse, $E_{i}\left( \omega\right)$, describes the amplitude and phase changes due to absorption and refraction of the traversed medium. This ratio is referred to as the material transfer function, $H\left( \omega\right)$. In the simple case of a homogeneous dielectric slab of thickness $d$ and complex refractive index $\hat{n}_s$, surrounded by nitrogen, the theoretical expression of the transfer function can be written, for normal incidence of waves, as~\cite{Withaya_14}:

\begin{align}
\label{Hfun1}
	H(\omega) & = \frac{E_t(\omega)}{E_i(\omega)} \nonumber \\
	& =  \tau\tau' \exp{\left\lbrace  -i\left[ \hat{n}_s(\omega)-n_0\right]  \frac{\omega d}{c}\right\rbrace } \cdot FP(\omega),
\end{align}

\begin{align}
\label{HfunFP}
	FP(\omega)& =\sum\limits_{m=0}^\infty \left\lbrace  \rho'^2 \exp{\left[-2i\hat{n}_s(\omega) \frac{\omega d}{c}\right]} \right\rbrace^m \nonumber\\
	&=\left\lbrace  1-\rho'^2 \exp{\left[-2i\hat{n}_s(\omega) \frac{\omega d}{c}\right]} \right\rbrace^{-1},
\end{align}
where 
\begin{equation}
\tau  = 2/(n_0+\hat{n}_s)
\end{equation}
is the nitrogen-sample complex transmission coefficient and 
\begin{equation}
\tau'=2\hat{n}_s/(n_0+\hat{n}_s), \rho' = (n_0-\hat{n}_s)/(n_0+\hat{n}_s)
\end{equation}
are the sample-nitrogen complex transmission and reflection coefficients, with $\hat{n}_s=n_s(\omega)-ik_s(\omega)$, where $n_s(\omega)$ is the refractive index, $k_s(\omega)$ the extinction coefficient, and $n_0$ the refractive index of nitrogen. 
Yet in eq.s~\ref{Hfun1} and \ref{HfunFP}, $c$ is the vacuum speed of light and $FP(\omega)$ represents the Fabry-P\'{e}rot effect due to the multiple reflections inside the sample. 
In a THz-TDS transmission experiment the optical properties of a material can be completely characterized by measuring the experimental transfer function, $H_{exp}( \omega)$, from which by using eq.s ~\ref{Hfun1} and \ref{HfunFP} the refractive index, $n_s\left( \omega\right)$, the absorption coefficient, $\alpha_s(\omega)=2\omega k_s(\omega)/c$ and the thickness could be in principle extracted. However, eq~\ref{Hfun1} is not in a closed form and cannot be solved to give analytical expressions for the optical parameters, moreover the thickness of the sample is not generally known with a sufficient accuracy.  An iterative process of calculation has to be employed~\cite{Withaya_05,Pupeza_07,Scheller_11,Scheller_09,Scheller_09b}. In our work we followed the numerical optimisation algorithm proposed by Scheller et all.~\cite{Scheller_11}. As a first step we can obtain raw estimations of $n_s$ and $\alpha_s$ by neglecting the $FP$ term and the imaginary part of the refractive index in the Fresnel coefficients of eq.~\ref{Hfun1}. With these approximations, analytical expressions for the optical parameters can be obtained~\cite{Withaya_14}:
\begin{align}
	n_s(\omega)&=n_0-\frac{c}{\omega d}\text{arg}\left[H(\omega)\right]   \label{ns}\\
	k_s(\omega)&=\frac{c}{\omega d}\left\lbrace ln \left[ \frac{4n_0n_s}{\vert H(\omega) \vert(n_0+n_s)^2}\right] \right\rbrace \label{alphas}
\end{align}

\begin{figure}[t]
\centering
\includegraphics[width=0.45\textwidth]{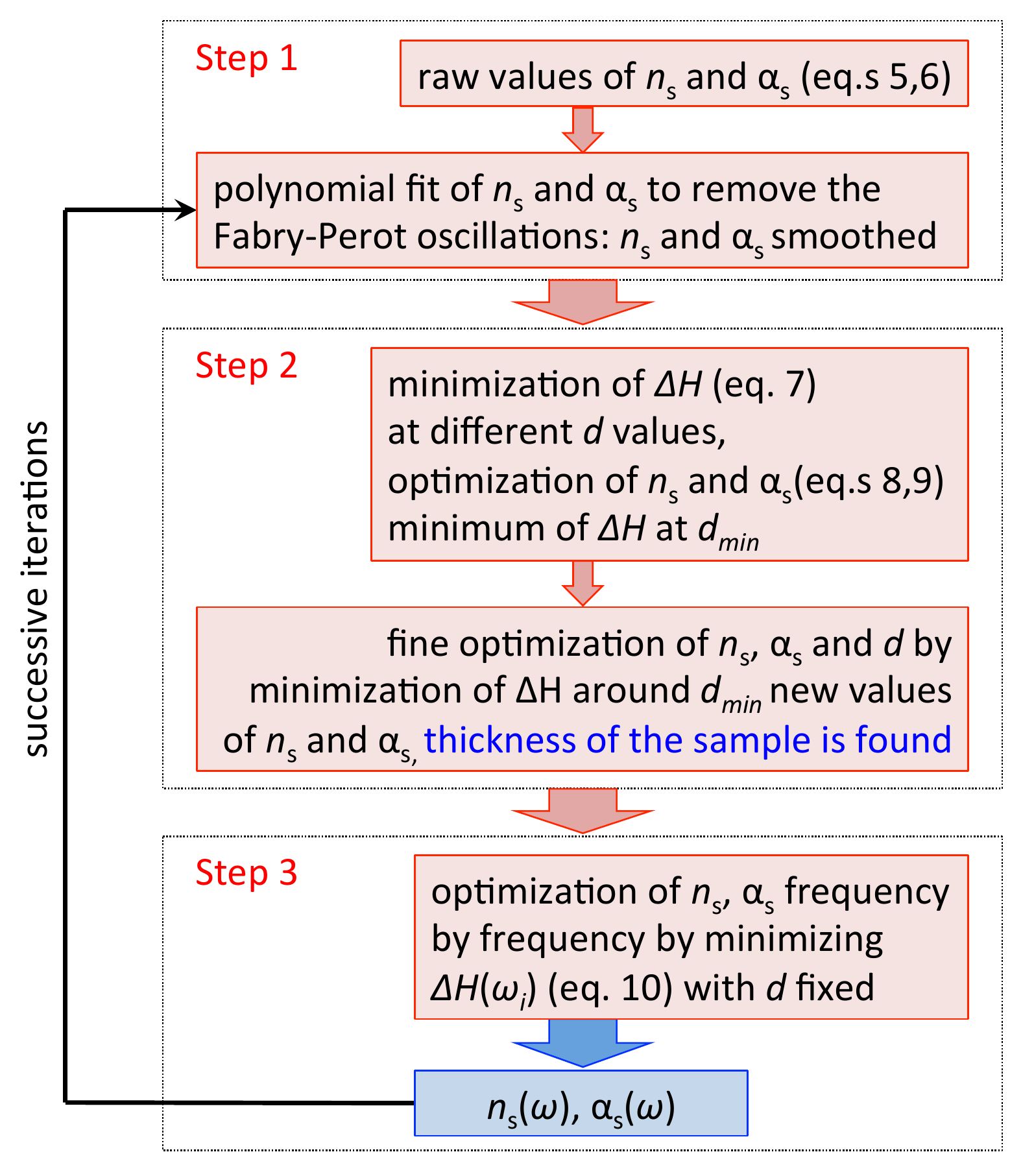}
\caption{Schematic representation of the fitting procedure used for the extraction of the refractive index, $n_S$, the absorption coefficient $\alpha_S$, and the sample thickness $d$. The algorithm can be diagrammed in three main blocks. Step 1: a preliminary and approximated evaluation of the refractive index and absorption coefficient. Step 2: a minimization routine of the function $\Delta H$ enables the estimation of the correct value of the sample thickness and more reliable values of  $n_S$ and $\alpha_S$. Finally, step 3: the real frequency dependence of the optical parameters is revealed through the minimization of $\Delta H$ frequency by frequency. For very thin samples successive iterations of the process need to be repeated several times as far as the sample thickness value is stabilized. In each cycle the output parameters of the step 3 are smoothed out by the polynomial fit and used as input parameters again for the step 1.}
\label{DiagBlock}
\end{figure}

Substituting in $H$ the experimental value $H_{exp}$ and assuming the initial value of $d$ what measured with a micrometric screw, we obtain approximated frequency dependent values of $n_s$ and $\alpha_s$, which are, moreover, affected by fake oscillations due to the neglected $FP$ effect. If the $FP$ reflection pulses are clearly distinguishable in the sample temporal signal, we can calculate the experimental transfer function with time shorten signals, where the reflection pulses have been simply cut off. This returns optical parameters not affected by the fake oscillations. However, when the reflections are close in time and partially superimposed, due to a short optical path, the cutting process can't be applied. Also in the case where the reflection peaks are well separated but the sample signal shows a long time evolution after the main peak, because of a structured absorption of the medium, the method can give wrong evaluations of the optical parameters. Anyway, it is better to use the full $H_{exp}$ and remove the oscillations in different way. Here we implement a polynomial fit of the optical parameters, of variable order and fitting range, by which we can catch the real physical frequency behaviour and remove the $FP$ oscillations.
After this preliminary evaluation of $n_s$, $\alpha_s$, and $d$, we can calculate the full theoretical expression of $H(\omega)$, eq.~\ref{Hfun1} together with eq. \ref{HfunFP}, with the summation of the $FP$ limited to the number of reflections appearing in the time window of the measurement. Then we can compare the result with the experimental one to infer new best values for $n_s$, $\alpha_s$, and $d$. Thus, the second step is to minimize the function
\begin{equation}\label{deltaH}
\bigtriangleup H=\sum\limits_{\omega}\vert H(\omega)-H_{exp}(\omega)\vert
\end{equation}
with a numerical optimization on the $n_s$, $\alpha_s$ for different fixed values of $d$. We use a Nelder-Mean simplex algorithm with the two scalars $\xi$ and $\psi$:
\begin{align}
n_{s,new}(\omega)&=\xi \left[ n_{s,old} (\omega)-1 \right]+1,\label{parn}\\
k_{s,new}(\omega)&=\psi k_{s,old}(\omega),\label{park}
\end{align}
For every value of $d$, new values of $n_s(\omega,d)$ and $\alpha_s(\omega,d)$ are calculated by eq.s~\ref{ns} and \ref{alphas}, filtered with the polynomial fit, and then optimized minimizing $\bigtriangleup H$. Graphing the minima of $\bigtriangleup H$ at varying of $d$ we plot a curve with a minimum for $d_{min}$ value, which corresponds to the thickness of the sample. The fitting process is then repeated, starting from the triad, $n_s(\omega,d_{min})$, $\alpha_s(\omega,d_{min})$, and $d_{min}$, but now with the additional parametrization of $d$,  $d=\zeta d$, in order to refine its value.
It is worth noting that the parametrizations of $n_s$ and $\alpha_s$ through the scalar $\xi$ and $\psi$ do not change their frequency behaviours which are still those inferred from the first step and can be affected by the filtering process.
Thus, as a third and final step, as already reported by Scheller et all.~\citep{Scheller_11}, we perform an optimization of the optical parameters at every frequency step $\omega_i$ using the function
\begin{equation}\label{deltaHwi}
\bigtriangleup H(\omega_i)=\vert H(\omega_i)-H_{exp}(\omega_i)\vert
\end{equation}
The starting values for $n_s$ and $\alpha_s$ are the optimal ones found in the previous step, the parametrizations are the same of eq.s~\ref{parn} and \ref{park} with the same algorithm, whilst $d$ is always kept fixed to the optimal value estimated before.
This last optimization reshapes frequency features of the optical constants that may have been distorted or erased by the first step evaluation and filtering process. 
This new set of curves of $n_s$ and $\alpha_s$ can be used for a new optimization cycle starting again with the step one: especially for a sample with short optical paths, the optimization must be repeated several times to find reliable values of the thickness and the optical constants. 
All the calculations and minimization routines written above by which all the data reported in this work have been processed, were performed by executing an in-house developed Matlab code. In Figure~\ref{DiagBlock} we report a block diagram of the fitting procedure.
What described so far concerns the analysis for a free standing single slab or layer, in the case of a bilayer system the optimization process is similar but starts from a different set of equations. The analysis can be carried out if at least the optical proprieties of one of the layers and its thickness are completely known. This can be achieved by preliminarily, characterizing one of the two layer as a single free standing layer by means of the analysis just now described. 
The first step is to consider the bilayer system as a single layer and obtain effective optical parameters using the approximated eq.s~\ref{ns} and \ref{alphas} with $d=d_1+d_2$, where $d_1$ and $d_2$ are the thicknesses of the two layers. The $n_{eff}$ and $\alpha_{eff}$ can be connected to the optical constants of the two layers by simple considerations on the refractive index and absorption. Denoting the layer under study as 1 and the known layer as 2, we get:
\begin{align}\label{n1k1}
n_1&=\frac{1}{d_1}\left[(n_{eff}-n_0)(d_1+d_2)-d_2(n_2-n_0)\right]+n_0,\\
k_1&=\frac{1}{d_1}k_{eff}(d_1+d_2)-k_2 \frac{d_2}{d_1},
\end{align}
The optical parameters calculated with these expressions need to be filtered from the $FP$ oscillations applying the same polynomial fit described before.  The transfer function for a bilayer system  now is more complex and, for waves at normal incidence, can be written as~\cite{MacFarlane_94,Jin_14}:
\begin{widetext}
\begin{align}\label{Hfun2}
H(\omega) &= \frac{E_t(\omega)}{E_i(\omega)} =  \frac
{\tau_{01}\tau_{12}\tau_{20}~e^{ -i\frac{\omega}{c}\left[  d_1\hat{n}_1+d_2\hat{n}_2-n_0\left( d_1+d_2\right) \right] } }
{\left[1-\rho_{21}\rho_{20}~e^{-i\frac{2\omega}{c}d_2\hat{n}_2}\right]
\left[ 1-\rho_{12}\rho_{10}~e^{-i\frac{2\omega}{c}d_1\hat{n}_1}-\frac{\rho_{20}\rho_{10}\tau_{21}\tau_{12}~e^{-i\frac{2\omega}{c}\left( d_1\hat{n}_1+d_2\hat{n}_2\right)}}{1-\rho_{21}\rho_{20}~e^{-i\frac{2\omega}{c}d_2\hat{n}_2}} \right]} 
\end{align}
\end{widetext}
where $\hat{n}_i$ are the complex refractive indices, $\tau_{ij}$ and $\rho_{ij}$ the complex transmission and reflection coefficients with $i,j=0$ for nitrogen, $1$ and $2$ for the two layers. This expression includes a $FP$ effect with an infinite number of reflexes between all the three separation surfaces; this can be considered as a valid approximation for thin layers and  measurements extended over long time delays. Following the same procedure of optimization and minimization as in the single slab case, but using the correct expression of $H(\omega)$ reported in eq.\ref{Hfun2} we can calculate the final values of the refractive index $n_1$ and the extinction coefficient $k_1$. 
\begin{figure*} [htb]
\centering
\includegraphics[width=0.7\textwidth]{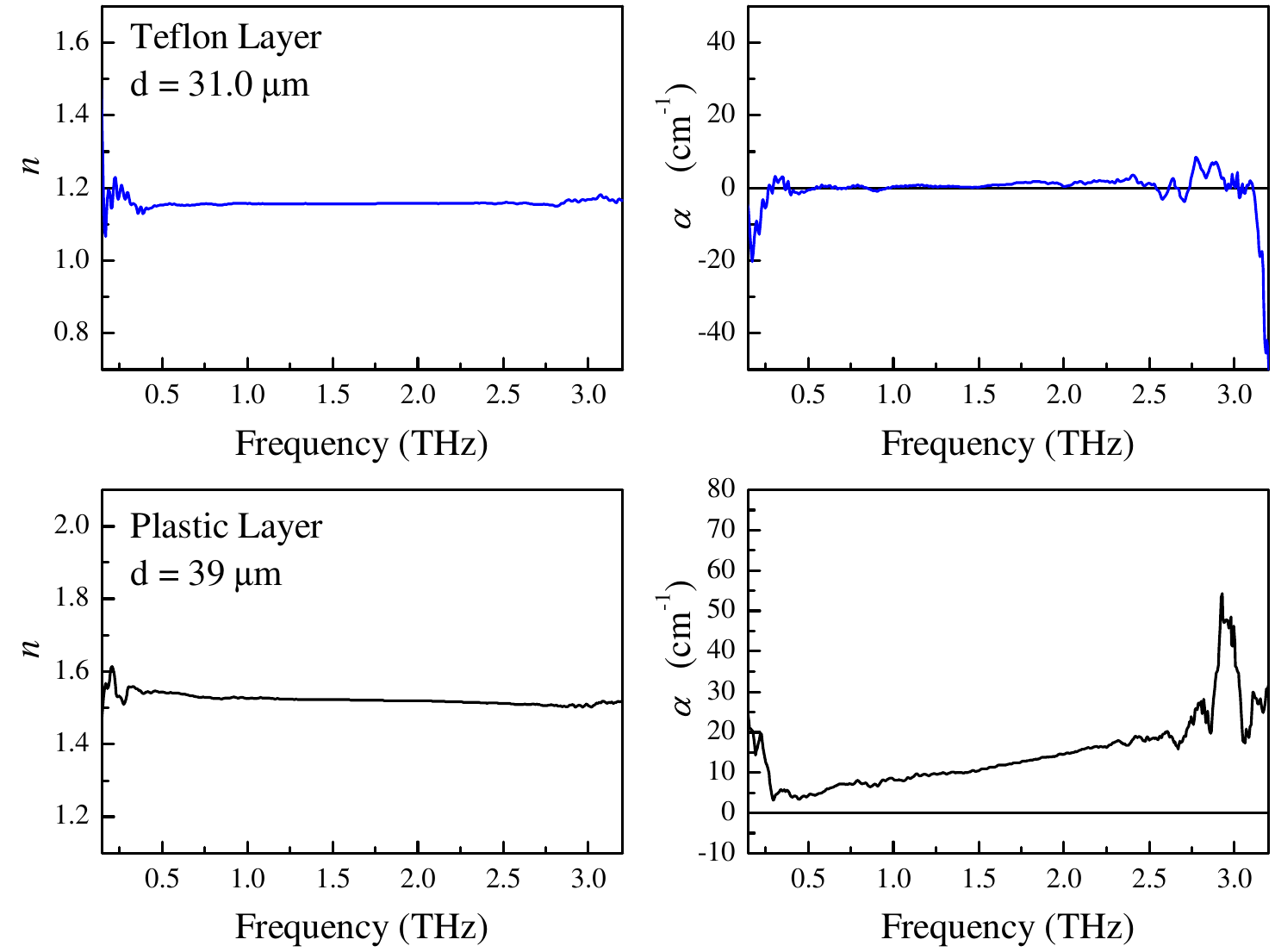}
\caption{Refractive index and absorption coefficient vs frequency of the Teflon layer and the plastic layer measured as single free standing layers. The analysis algorithm for a single layer sample gives a thickness of $31~\mu m$ for the Teflon layer and $39~\mu m$ for the plastic layer.}
\label{singlelayer}
\end{figure*}

\section{Results and discussion}
In order to in-depth test the analysis algorithm for the extraction of the optical parameters and thickness in single and bilayer samples, we studied three samples; a single layer made of Teflon, a single plastic layer and a bilayer sample made by the close overlap of these two layers. The single layer thicknesses measured by a micrometric screw equal $30\pm 2~\mu m$ and $37\pm 2~\mu m$, respectively\footnote{Due to the softness of the two materials, we had to hold the layers between two glass windows and calculate the layer thickness as a difference.}. The single layer samples have been studied as free standing samples and then glued to form a bilayer system (the glue is estimated to have sub-micrometric thickness). The Teflon layer is made mainly of politetrafluoroetilene (PTFE); the plastic layer is made mainly of polipropilene (PP). Nevertheless, some other polymers could be present in these material compositions, hence the absolute values of their indexes of refraction in the THz range are unknown.
In Figure~\ref{singlelayer} we show the results obtained for the two materials individually studied. The single layer analysis gave for the Teflon sample a refractive index almost constant in all the probed frequency range, $n=1.18$, and a negligible absorption coefficient. For the plastic layer, instead, we found a refractive index weakly decreasing with the frequency, with a value of $n=1.53$ at $1~THz$, and an absorption coefficient increasing with the frequency ($\alpha\sim10~cm^{-1}$ at $1~THz$).

\begin{figure*}
\centering
\includegraphics[width=0.7\textwidth]{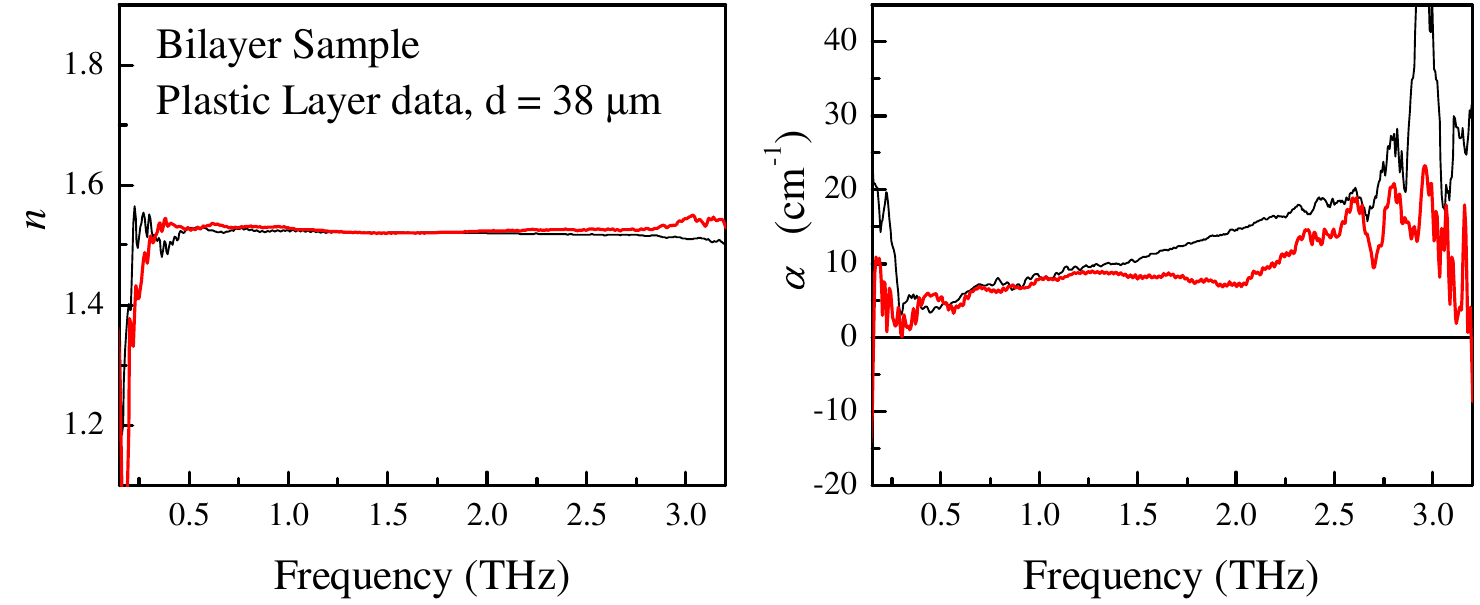}
\caption{Refractive index and absorption coefficient vs frequency of the plastic layer measured in the bilayer configuration (red line) and extracted from the bilayer analysis considering the Teflon layer as known material. For comparison, we report in the figure the plastic layer measured as single free standing sample (black line). The optical parameters are compared to those obtained from the single layer measurement with a very good agreement.}
\label{bilayer}
\end{figure*}

The measured values are in fair agreement with other THz measurements on PTFE and PP bulk samples \cite{Jin_06}. 
The sample thickness, extracted from the single layer analysis, for the Teflon and the plastic samples were $31~\mu m$ and $39~\mu m$, respectively. These values are in a good agreement with the values measured with the micrometric screw. As a second step in the test, we performed the analysis of the bilayer system considering as known material the Teflon layer and the unknown one the plastic layer;  we extracted the thickness and optical parameters of the plastic layer by using the bilayer analysis algorithm and compared the results with those extracted by the single layer analysis done before. Figure ~\ref{bilayer} reports the comparison between the two analysis. The agreement of thickness, refractive index, and absorption coefficient is really good. The absolute error we estimate on the thickness extractions is of about $5~\mu m$ for both the analysis. Moreover, the data analysis has a lower limit of about $10~\mu m$ for thickness for a trustworthy measurement of the thickness values.  
\begin{figure*} 
\centering
\includegraphics[width=0.7\textwidth]{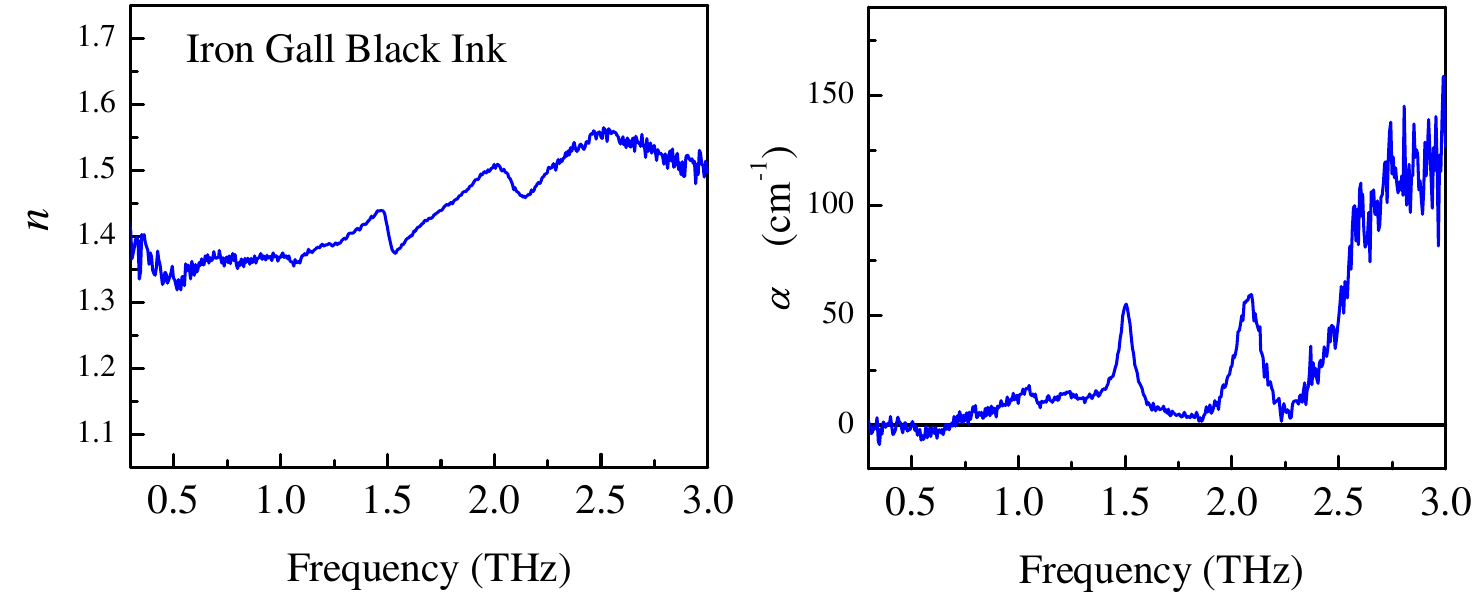}
\caption{THz behaviour of the refractive index and absorption coefficient of iron gall black ink studied in the form of thin film layered on $10~\mu m$ PE pellicle. The bilayer algorithm allows for the measurement of the ink film thickness of about $17~\mu m$.}
\label{bilayerink}
\end{figure*}
After this hard test of the analysis procedure and in order to get closer to a real-practice experiment on ancient manuscripts and drawings, we measured a black ink film layered on polyethylene (PE) pellicle of $10~\mu m$ (polyethylene far-IR sample cards by Sigma-Aldrich). As an ink sample we chose an historical iron gall black ink prepared in lab using as a source of gallo-tannic acid a synthesized gallic acid (more details about the ink preparation and the comparison with other black inks can be found on~\cite{tasseva_17}). PE is the ideal support for absorption spectroscopy in the THz region thanks to its negligible absorption coefficient (below of $1~cm^{-1}$ see~\cite{Lee_09} and references therein) and a refractive index equal to 1.4 constant in all the studied THz frequency range.
In Figure \ref{bilayerink} we report the optical parameters for the studied ink. A thickness of about $17~\mu m$ was found. The ink shows an absorption spectra with features that have to be ascribed to absorption of the gallic acid present in the ink \cite{taschin_17}. These features are confirmed by the dispersive character of the refractive index at these frequencies.

\section{Conclusion}
We implemented an innovative experimental procedure and data analysis to measure the transmission parameters in the THz frequency range and the thickness of thin film materials. 
In the data analysis we used a new method in order to extract the material characteristics from the THz-TDS data. In the case of thin film materials, the THz pulse transmission is strongly affected by multiple reflections. For these samples, the extraction of the real physical material parameters requires a proper data analysis taking into account the multiple reflection contributions to the THz-TDS signal. We implemented an iterative fitting process based on a polynomial fit of the transmission parameters that enables a correct extraction of the refraction indexes and absorption coefficients for samples with thickness down to $10~\mu m$, both free standing layers and multi-layer system.

Using the THz-TDS technique and the iterative fitting procedure of the data, we succeed to measure with high confidence the refraction indexes and the absorption coefficients of samples made of a single thin layer or double layer structure. The study shows that the frequency dependence of the transmission parameters of very thin layered samples can be extracted reliably from the THz-TDS measurements, disentangling the single layer contributions. Moreover, the THz transmission parameters of each layer are measured in absolute scale of values and the layer thickness is extracted.

We applied this techniques to the following samples: a couple of samples made of teflon and polypropylene, with single layer and bilayer structures, in order to test the experimental and data analysis potentialities; A prototype sample for artworks investigation made by a thin film of black ink layered on a polyethylene pellicle. 

\section*{Acknowledgement}
This work was founded by Regione Toscana, prog. POR-CROFSE-UNIFI-26 and by Ente Cassa di Risparmio Firenze, prog. 2015-0857. We acknowledge M. De Pas, A. Montori, and M. Giuntini for providing their continuous assistance in the electronic set-ups; and R. Ballerini and A. Hajeb for the mechanical realizations.

\section*{References}
%

\end{document}